\documentclass[review,12pt,3p]{elsarticle}

\usepackage{amsmath}
\usepackage{amssymb}
\usepackage{verbatim}  
\usepackage{siunitx}
\usepackage{algorithmic}
\usepackage{algorithm}
\usepackage{array}
\usepackage{eqparbox}

\begin{document}

\begin{frontmatter}

\title{Distributed Cooperative Spectrum Sharing in UAV Networks Using Multi-Agent Reinforcement Learning}

\author[label1]{Alireza Shamsoshoara}
\address[label1]{School of Informatics, Computing and Cyber Systems, Northern Arizona University, Flagstaff, AZ, USA}

\author[label1]{Mehrdad Khaledi}

\author[label1]{Fatemeh Afghah}

\author[label1]{Abolfazl Razi}

\author[label2]{Jonathan Ashdown}
\address[label2]{Computer Information Systems Department, SUNY Polytechnic Institute, Utica, NY, USA}

\begin{abstract}
In this paper, we develop a distributed mechanism for spectrum sharing among a network of unmanned aerial vehicles (UAV) and  licensed terrestrial networks. This method can provide a practical solution for situations where the UAV network may need external spectrum when dealing with congested spectrum or need to change its operational frequency due to security threats. Here we study a scenario where the UAV network performs a remote sensing mission.   In this model, the UAVs are categorized to two clusters of relaying and sensing UAVs. The relay UAVs provide a relaying service for a licensed network to obtain spectrum access for the rest of UAVs that perform the sensing task. We develop a distributed mechanism in which the UAVs locally decide whether they need to participate in relaying or sensing considering the fact that communications among UAVs may not be feasible or reliable. 
The UAVs learn the optimal task allocation using a distributed reinforcement learning algorithm. Convergence of the algorithm is discussed and simulation results are presented for different scenarios to verify the convergence\footnote{This material is based upon the work supported by the National Science Foundation under Grant No. 1755984.}. 
\end{abstract}


\end{frontmatter}



\section{Introduction}
Unmanned Aerial Vehicles (UAVs) have been recently used in many civilian, commercial and military applications \cite{Razi_Asilomar17,Peng2018,Mousavi_INFOCOM18,Afghah_ACC18,Khaledi_SECON18,Afghah_NWRCS}. With recent advances in design and production of UAVs, the global market revenue of UAVs is expected to reach \$11.2 billion by 2020 \cite{Gartner_UAV}. 

Spectrum management is one of the key challenges in UAV networks, since spectrum shortage 
can impede the operation of 
these networks. In particular, in applications involving a low-latency video streaming, the UAVs may require additional spectrum to complete their mission. The conventional spectrum sharing mechanism such as spectrum sensing may not be very practical in UAV systems noting the considerable required energy for spectrum sensing or the fact that they cannot guarantee a continuous spectrum access. The property-right spectrum sharing techniques operate based on an agreement between the licensed and unlicensed users where the spectrum owners lease their spectrum to the unlicensed ones in exchange for certain services such as cooperative relaying or energy harvesting.

In this paper, we studied the problem of limited spectrum in UAV networks and considered a relay-based cooperative spectrum leasing scenario in which a group of UAVs in the network cooperatively forward data packet for a 
ground primary user (PU) in exchange for spectrum access. The rest of the UAVs in the network utilize the obtained spectrum for transmission and completion of the remote sensing operation. Thus, the main problem is to partition the UAV network into two task groups in a distributed way.

It is worth noting that cooperative spectrum sharing has been studied previously in the context of cognitive radio networks \cite{Stanojev08,Afghah_CDC2013,Korenda_CISS,Namvar15}. The existing models are mostly centralized and the set of relay nodes 
is typically chosen by the PU. Such solutions, however, are not applicable to UAV networks, due to their distributed 
infrastructure and autonomous functionality. 

To tackle this problem, we utilize multi-agent reinforcement learning \cite{Guestrin2002,Lauer2000,mousavi2016deep,Chalkiadakis2003, mousavi2017traffic, mousavi2016learning, mousavi2017applying}, which is an effective tool for designing algorithms in distributed systems, where the environment is unknown and a reliable communication among agents is not guaranteed. The main problems in distributed multi-agent reinforcement learning 
include dealing with state space complexity and 
the lack of complete information about other agents. There have been proposals in the literature to address these issues through message passing or simplifying assumptions. For instance, \cite{Guestrin2002} assumes that the decision of an agent depends only on a limited group of other agents, which decomposes the state space and simplifies the problem. In another work \cite{Chalkiadakis2003}, a Bayesian setting is proposed where each agent has some distributional knowledge about other agents' decisions. Such simplifications, however, are not applicable to the distributed UAV network environment.

In this paper, we propose a distributed multi-agent reinforcement learning algorithm for task allocation among UAVs. Each UAV either joins a relaying group to provide relaying service for the PU or performs data transmission to the UAV fusion center. In this approach, each UAV maintains a local table about the 
respective rewards for
its actions in different states. The tables are updated locally based on a feedback from PU receiver and the UAV fusion node. We define utilities for both the PU and the UAV network, and the objective is to maximize the total utility of the system (sum utility of the PU and the UAV network). We discuss the convergence of our learning algorithm and we present simulation results to verify the convergence to the optimal solution.

The remainder of this paper is organized as follows. In Section \ref{sec:SystemModel}, the system model and the assumptions of the proposed model are described. In Section \ref{sec:ProposedMethod}, we propose a distributed multi-agent learning algorithm to solve the spectrum sharing problem.  In Section \ref{sec:Simulations}, we present simulation results and discuss the performance of our distributed learning algorithm. Finally, we make concluding remarks in Section \ref{sec:Conclusion}.

\section{System Model}\label{sec:SystemModel} 
\begin{figure}[t]
\centering
\includegraphics[width=0.7\linewidth]{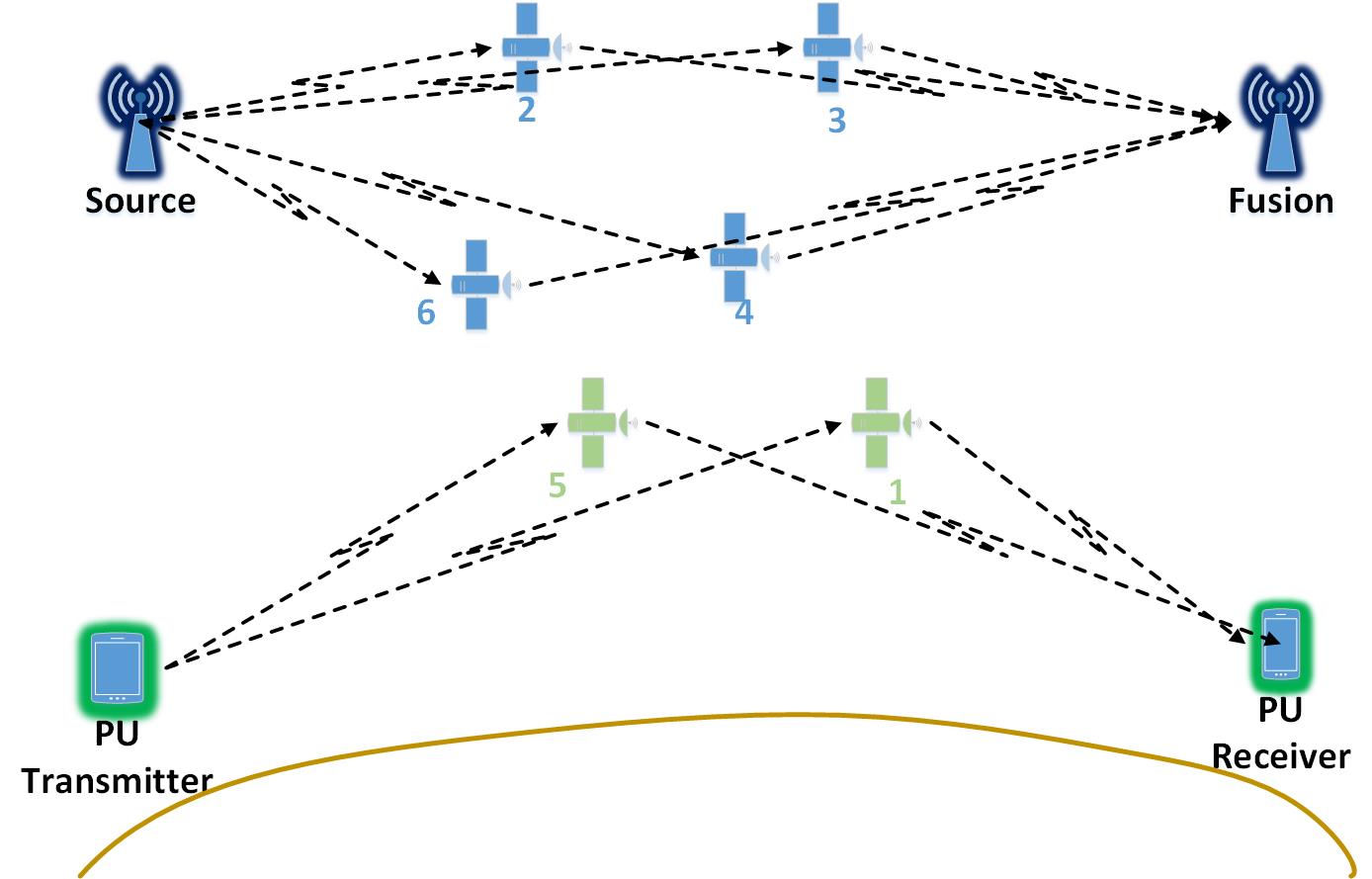}
\caption{System Model: A sample Scenario with 6 UAVs, where four UAVs handle  packets relaying between the Source and Fusion Center and two UAVs relay packets for the Primary User. }
	\vspace{-10pt}
\label{fig:System Model}
\end{figure}

We consider a licensed primary user (PU) who is willing to share a part of its spectrum with a network of UAVs, in exchange for receiving a cooperative relaying service. The UAV network consists of $N$ UAVs which can be partitioned into two sets depending on the task of the UAV. In fact, UAVs either relay for the PU or utilize the spectrum to transmit their own packets to the fusion center. Let $K$ be the number of nodes who perform the relaying task and $N - K$ denote the number of  UAVs that transmit packets to their fusion center. In this paper, we assume that both the PU's transmitter and receiver are terrestrial, while UAVs are operating in high elevation. Also, we assume no reliable direct link exists between the PU's transmitter and receiver. Moreover, there is zero chance for direct transmission between the UAVs' source and fusion, due to their distances from the fusion center. Fig. \ref{fig:System Model} illustrates a sample scenario with 6 total nodes, where 
the nodes are partitioned into a set of 4 
relay nodes for the fusion center, and 2 other nodes relay information for the PU receiver on the ground.

The PU's transmitter intends to send its packet to a designated 
receiver, which is far away from its location. Hence, a single or a number of UAVs are required to deliver its information to the receiver. In addition, we assume that the UAVs' spectrum is  congested or unreliable, therefore the UAVs  are required to lease additional spectrum from the PU to communicate with their fusion. 
By delivering the PU's packet, the UAVs  gain  spectrum access to send their own packets. All the UAVs transmitters and receivers are assumed to be equipped with a single antenna. Also, we assume that the channels between UAVs, source, fusion, and PU transmitter and receiver are slow Rayleigh fading with a constant coefficient over one time slot. 
The channel coefficients are defined as follows: i) $h_{PT,U_i}$ refers to the channel parameters between the PU's transmitter and $i^{th}$ UAV; ii) $h_{U_i,PR}$ denotes the parameters between the $i^{th}$ UAV and the PU's receiver; iii) $h_{S,U_i}$ and $h_{U_i,F}$, respectively denote the channel coefficients between the Source and the $i^{th}$ UAV, and between the $i^{th}$ UAV and the fusion center. 
For the sake of simplicity, the instant Channel State Information (CSI) are assumed to be available for all UAVs 
following similar works in \cite{stanojev2013improving,wu2011information,al2016enhancingg, afghah2018reputation,shamsoshoara2015enhanced}

\begin{figure}[t]
\centering
\includegraphics[width=0.7\linewidth]{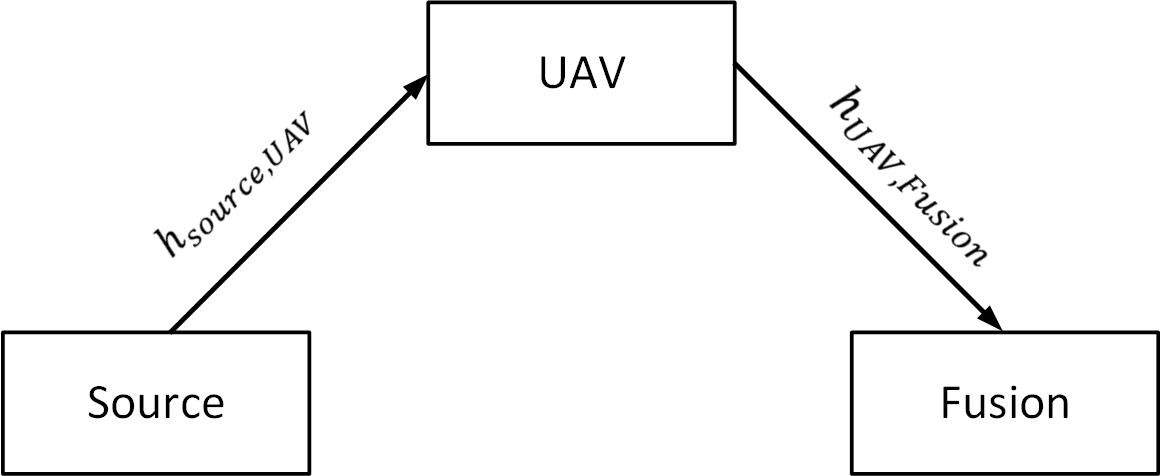}
\caption{Communication channels for a single relay}
	\vspace{-10pt}
\label{fig:Sinle Relay model}
\end{figure}

The source of the noise at the receivers 
is considered as 
a symmetric \textit{normally} distributed random variable, denoted by 
$z \sim CN(0, \sigma^2)$. Many works such as \cite{roberge2013comparison, mozaffari2016optimal,shamsoshoara2015enhanced} optimized the power consumption and nodes' lifetime in this area. On the other hand, power optimization is not the purpose of this work, hence we assume constant powers during the transmissions. However, the transmission power for the Source and the PU transmitter is less than
those of the UAVs. Half-duplex strategy is utilized in this work. Without loss of generality, time-division notations are characterized in order to ensure the half-duplex operations. After these assumptions, the channel and system model for a single relay is shown in Fig. \ref{fig:Sinle Relay model}. In this model, all UAVs and terminals utilize a single antenna for transmission.

In the first half of a transmission 
cycle, 
the source transmits its packet and 
the relay UAVs receive the information.
The channel model for the first half is presented as follows: 
\begin{align}\label{eq:1}
y_{U}[n] = h_{S, U}x_{S}[n] + z_{r}[n],	
\end{align}
where $x_{S}$ is the source's transmitted signal and $y_{U}$ is the UAV's received signal. Then, in the second half of the transmission, the UAV sends the received packet 
in the previous time slot. 
We can write the second half as another model for the received signal as follow:
\begin{align}\label{eq:2}
y_{F}[n] = h_{U, F}x_{U}[n] + z_{F}[n],	
\end{align}
where $x_{U}$ is the UAV's transmitted signal and $y_{F}$ is the destination's received signal. 

\begin{figure}[t]
\centering
\includegraphics[height=5in, width=0.5\linewidth]{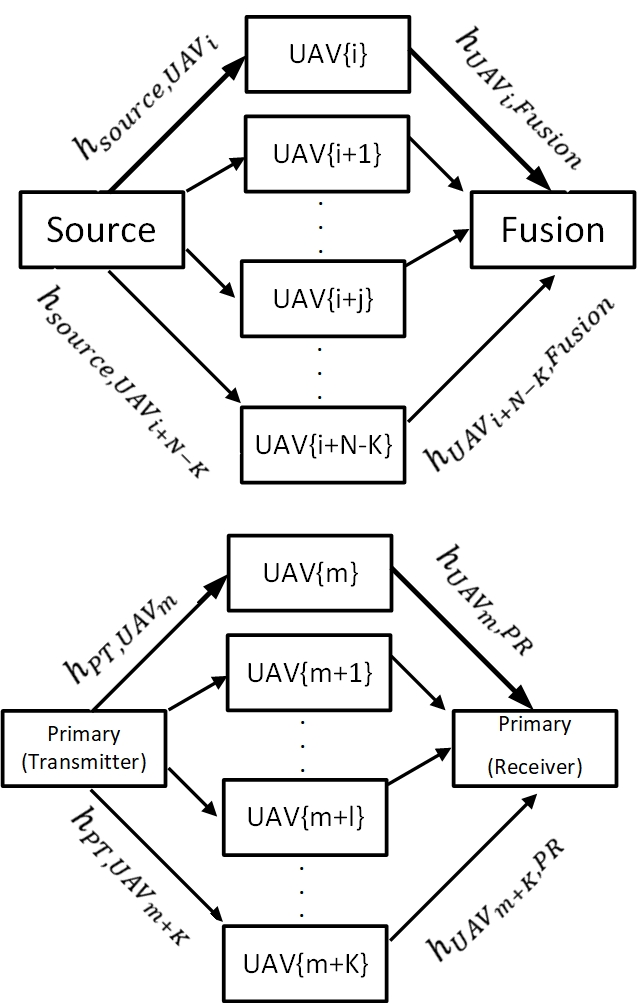}
\caption{System Model: Dividing UAVs into $K$ and $N-K$ groups, for cooperating in two sets of Source-Fusion and Primary Transmitter-Receiver.}
	\vspace{-10pt}
\label{fig:Groups}
\end{figure}

In equations (\ref{eq:1}) and (\ref{eq:2}), the CSI parameters $h_{ij}$ represent the effects of the path loss and likewise $z_{j}$ represents the effect of noise and interference terms at the receiver, where $i \in \{\textnormal{Source}, \textnormal{PU-Transmitter}, \textnormal{UAV}\}$ and $j \in \{\textnormal{Fusion}, \textnormal{PU-R}, \textnormal{UAV} \}$. In our scenario, $h_{ij}$ is calculated by the proper receiver. 
\\
Based on equations (\ref{eq:1}) and (\ref{eq:2}), the throughput capacity of the non-degraded discrete memoryless broadcast channel is expressed in (\ref{eq:3}) \cite{AnalyzingAFZhang}: 
\begin{align}\label{eq:3}
C_{Throughput} = \max_{w \rightarrow x \rightarrow y_{d}} \{I(w;y_{d})\},
\end{align}
where $d \in \{\textnormal{Fusion}, \textnormal{PU-Receiver}\}$, $w$ is the message word and $x$ is the codeword which has been assigned to each message by the encoder. Preferably, equation (\ref{eq:3}) should be solved for the optimal joint distribution of both $w$ and $x$. 
However, as discussed in \cite{shafiee2007achievable}, we can achieve the suboptimal throughput rate in (\ref{eq:4}), with the aid of assumption $x = w$. Also, $p(x)$ denotes the probability mass function (pmf) for the codeword.
\begin{align}\label{eq:4}
R_{Throughput} = \max_{p(x)} \{I(x;y_d)\}
\end{align}
In scenarios, where 
users can exploit the existence of UAVs, different cooperation protocols such as Decode and Forward (DF) and Amplify and Forward (AF) can be used \cite{laneman2001efficient}. 
The idea behind the concept of \textit{cooperative relaying} is that a set of relay nodes decode, amplify and collectively ``beam-form" the signal received from the source node (potentially with help of source node itself) towards a designated destination in order to exploit transmission diversity and increase the overall throughput of the system \cite{levin2012amplify}. 
\\
Considering an AF cooperation, 
each  UAV first amplifies the signals from the source and then cooperates with source to send its information to the fusion center or to the PU-Receiver. According to \cite{laneman2004cooperative}, the mutual information for i) the first set of source, UAV, fusion and ii) the second set of PU-T, UAV, PU-R can be written as equations (\ref{eq:5}) and (\ref{eq:6}), respectively. In these equations, $P_S$ denotes the transmitter power from the source of the UAV network and $i$ specifies the index for the UAV.
\begin{align}\label{eq:5}
&I_{SF_{AF}} = \log_2(1 + P_{S}|h_{SF}|^2  \\
\nonumber
&+ \frac{P_{S}|h_{S,U_i}|^2\: P_{U_i}|h_{U_i, F}|^2}{1 +P_{S}|h_{S,U_i}|^2 + P_{U_i}|h_{U_i, F}|^2})
\end{align}
\begin{align}\label{eq:6}
&I_{PU(TR)_{AF}} = \log_2(1 + P_{PT}|h_{PT,PR}|^2  \\
\nonumber
&+ \frac{P_{PT}|h_{PT,U_i}|^2 \: P_{U_i}|h_{U_i, PR}|^2}{1 +P_{PT}|h_{PT,U_i}|^2 + P_{U_i}|h_{U_i, PR}|^2})
\end{align}
We denote the throughput rate for both primary users and source-fusion users as (\ref{eq:7}) and (\ref{eq:8}), respectively.
\begin{align}\label{eq:7}
R_{PU} = I_{PU(TR)_{AF}} 
\end{align}
\begin{align}\label{eq:8}
R_{SF} = I_{SF_{AF}} 
\end{align}
It is noteworthy 
that these equations 
are valid only 
for cooperation with a single Relay or UAV. However, the 
objective 
of this paper is dealing with Multi-UAV or Multi-Agent relays. Fig. \ref{fig:Groups} demonstrates the distribution of $N$ UAVs into two groups 
including $K$ UAVs facilitating the air source-to-fusion communication and $N - K$ UAVs providing relaying service for a ground-based primary transmitter-receiver pair.
Hence, the equations for multi-UAV should be changed to (\ref{eq:9}) and (\ref{eq:10}). In (\ref{eq:9}), $i$ defines the lower bound for the first UAV in the source-fusion pair and $i + N - K$ denotes the upper bound. 
\begin{align}\label{eq:9}
&R_{SF}\textnormal{(Multi-UAV)} = \log_2(1 + P_{S}|h_{SF}|^2  \\
\nonumber
&+ \sum_{j = i}^{i+N-K} \frac{P_{S}|h_{S,U_j}|^2 \: P_{U_j}|h_{U_j, F}|^2}{1 +P_{S}|h_{S,U_j}|^2 + P_{U_j}|h_{U_j, F}|^2})
\end{align}
Here, $R_{SF}\textnormal{(Multi-UAV)}$ is the achievable rate 
for the fusion center.
This rate is achieved with the help of ($N-K$) UAVs. 
$P_S$ and $P_{U_i}$ are transmission powers for the source and the $i^{th}$ UAV, respectively. Also, $h_{SF}$ denotes the channel coefficient for the pair of source-fusion center, $h_{S,U_j}$ stands for the channel between the source and $j^{th}$ UAV, and finally $h_{U_j, F}$ denotes CSI for the $j^{th}$ UAV and the fusion. In (\ref{eq:10}), $m$ and $m + K$ define the lower and upper bound for the first  and last UAV in the source-fusion pair respectively.
\begin{align}\label{eq:10}
&R_{PU}\textnormal{(Multi-UAV)} = \log_2(1 + P_{PT}|h_{PT,PR}|^2  \\
\nonumber
&+ \sum_{l = m}^{m + K} \frac{P_{PT}|h_{PT,U_l}|^2 \: P_{U_l}|h_{U_l, PR}|^2}{1 +P_{PT}|h_{PT,U_l}|^2 + P_{U_l}|h_{U_l, PR}|^2})
\end{align}
In (\ref{eq:10}), $R_{PU}\textnormal{(Multi-UAV)}$ is the achievable rate for the primary transmitter-receiver pair with the aid of $K$ UAVs. 
$P_{PT}$ and $P_{U_l}$ are transmission power for the primary user and the $i^{th}$ UAV, respectively. Moreover, $h_{PT,PR}$, denotes the channel coefficients for primary transmitter and receiver. $h_{PT,U_l}$ stands for the primary transmitter and $l^{th}$ UAV. Finally $h_{U_l, PR}$ is CSI parameters for the $l^{th}$ UAV and the primary receiver.
Based on the assumption of long distance between the source and the fusion center and also the long distance between the primary transmitter and receiver, we can assume that $h_{SF}$ and $h_{PT,PR}$ are negligible.

Time is slotted and at the end of each time slot, the fusion center and the primary receiver send feedback to the UAVs informing them about the achieved accumulated rates. This information is used by each UAV to decide on joining a task group. The goal is to find the optimal task allocation for UAVs in a fully distributed way such that the total utility of the system (i.e. sum utility of UAV network (\ref{eq:9}) and the PU (\ref{eq:10})) is maximized. We assume that the UAVs decide locally with no information exchange among themselves. 

It is noteworthy that in some cases, 
the maximum throughput is achieved 
when all UAVs join the same set and deliver packets only for one set, which is not consistent with the proposed model. 
If all UAVs are distributed in the set of source-fusion, then the total throughput rate is zero because there is no available spectrum for UAVs to utilize for their transmission. Also, if all UAVs are partitioned in the primary set, then the sum throughput rate is equal to the rate of the primary user. In this case the proposed method handles this issue by considering the Jain fairness index \cite{lan2010axiomatic}. 
Based on the fact that we only have two sets and based on the Jain index definition, (\ref{eq:11}) describes the fairness for the proposed method in our system model. 
\begin{align}\label{eq:11}
J(x) = \frac{1}{n} \times \frac{(\sum_{i} x_{i})^2}{\sum_{i} x_{i}^{2}},
\end{align}

\noindent Here, $n$ is equal to 2 and $i \in \{0,1\}$ which indicates the set of source-fusion or Primary Users. We assume that $x_0$ and $x_1$ are equal to the number of UAVs in the Fusion-Source set and the Primary Users set, respectively. Therefore, we can define the fairness as (\ref{eq:12}).
\begin{align}\label{eq:12}
Fariness = \frac{1}{2} \times \frac{(\#\textnormal{U}_{F} + \#\textnormal{U}_{P})^2}{(\#\textnormal{U}_{F})^2 + (\#\textnormal{U}_{P})^2}
\end{align}
Now, if all UAVs are distributed in one set, then the fairness will be minimum ($0.5$), and if the UAVs are partitioned equally among two sets, then the fairness will be maximum ($1$). 

Based on these definitions, we define (\ref{eq:13}), as the gain value for each time slot which indicates the efficiency and performance for the distributed UAVs in two sets. 
\begin{align}\label{eq:13}
& Gain = \gamma_{1} \times \Delta(Rate_{Fusion})\\
\nonumber
&+\gamma_{2} \times \Delta(Rate_{Primary}) + \gamma_{3} \times (Fairness)
\end{align}

In (\ref{eq:13}), $\Delta(\textnormal{Rate}_{\textnormal{Fusion}})$ is the difference between the rate at time $t$ and the average of previous rates for the fusion center and $\Delta\textnormal{(Rate}_{\textnormal{Primary}})$ is the difference between the rate at time $t$ and the average of previous rates for the primary user. Also, $\gamma_{1}$, $\gamma_{2}$, and $\gamma_{3}$ are defined to control the gain value. Then, we use this gain in our proposed method as described in section \ref{sec:ProposedMethod}.

\color{black}
\section{The Distributed Learning Algorithm for Task Allocation}\label{sec:ProposedMethod}  



The proposed method is a general form of the Q-learning algorithm \cite{watkins1992q} for a distributed multi-agent environment.

Let $a^{i}_{t}$ denote the action chosen by UAV $i$ at time $t$, and let $A^i$ denote the set of all possible actions for UAV $i$. We consider two possible actions for a UAV that correspond to either joining the relaying task group or the fusion task partition. Therefore, the set of possible actions are identical across UAVs. We denote the action vector of UAVs at time $t$ by $u_t=(a^{1}_{t}, a^{2}_{t},\cdots, a^{N}_{t})$, and we refer to the set of all possible action vectors by $\mathcal{U}$. There is a finite set of states $\mathcal{S}$, where state $s\in \mathcal{S}$ corresponds to the current task partition. A deterministic transition rule $\delta$ governs the transition between states, i.e. $\delta: \mathcal{S}\times \mathcal{U} \rightarrow \mathcal{S}$. The reward function $r$ maps the current state and action vector to a real value, that is $r: \mathcal{S}\times \mathcal{U} \rightarrow \mathbb{R}$.
At the beginning of each time step, the UAVs observe the current state (this information is obtained by the feedback from the previous step). Then, each UAV independently decides on its action (i.e. which task group to join) without knowing any information about actions of the other agents. The rewards associated with the UAVs' actions are computed 
by the PU receiver and the UAV fusion. The reward is basically the gain obtained from the task partitioning, taking into account the utilities of the PU and the UAV network. After the reward is calculated, a feedback message from the PU receiver and the UAV fusion is broadcasted to the UAVs. This feedback message contains the reward and the current task partitions. 

The feedback information is used to update and maintain local Q-tables at each UAV. A Q-table basically represents the quality of different actions for a given state. For instance, $q^{i}_{t}(s,a)$ denotes the quality of action $a$ at state $s$ for UAV $i$ at time $t$. Individual Q-tables are updated as follows. At first, the tables are initialized with $q^{i}_{0} (s,a) = 0 $. Then, the following equation is used to update the Q-tables:

\begin{align} \label{eq:qUpdates}
q^{i}_{t+1} (s,a) = 
\begin{cases}
q^{i}_{t} (s,a),  \qquad\qquad\qquad \textnormal{if} \;s\neq s_t\; \textnormal{or} \;\;a\neq a^{i}_{t},  \\
\\
(1-\alpha) \:q^{i}_{t} (s,a) +\\
\qquad \alpha \cdot \big(r_{t} +\beta \cdot \max_{a'\in A^i} q^{i}_{t} (\delta(s_{t},u_t), a') \big), \\
\qquad\qquad\qquad\qquad\qquad \textnormal{otherwise},
\end{cases}
\end{align}
\noindent where $0\leq \alpha <1$ is the learning rate, $r_{t}$ is the reward or the gain obtained at time $t$, as defined in the system model, and $0\leq \beta <1$ is the discount factor to control the weight of future rewards in the current decisions.

The main idea is that in our distributed environment, the UAVs are unable to keep a global Q-table, 
corresponding to the 
current action vectors, i.e. $Q: \mathcal{S}\times \mathcal{U} \rightarrow \mathbb{R}$. Instead, each UAV $i$ keeps a local (and considerably smaller) Q-table which cares about its own current action, i.e. $q^i: \mathcal{S}\times A^i \rightarrow \mathbb{R}$. This approach significantly reduces the complexity of the algorithm and eliminates the need for coordination (or sharing information) with other UAVs at the time of decision making. However, we need a projection method that compresses the information of the global Q-table into the local small tables. 

The results in \cite{Lauer2000} prove that in a deterministic multi-agent Markov decision process and for the same sequence of states and actions, if every independent learner chooses locally optimal actions, the result would be the same as choosing the optimal action from a global table. We utilize this result and consider an optimistic projection method that assumes each UAV chooses the maximum quality action from its local table. This reasonable assumption is a necessary condition for the optimality of the learning algorithm. It is worth noting that the existence of a unique optimal solution is the sufficient condition for the optimality of this algorithm. It means that there should be a unique task partition, which results in the maximum total utility. If multiple task partitions yield the maximum utility, it is possible that the UAVs act optimally and choose the optimal actions in their local Q-tables, but the combination of their actions may not be optimal. In this case, message passing among UAVs is needed as they need to coordinate decisions at every step.

It should also be noted that in learning algorithms we need a balance between exploring new actions and exploiting the previously learned quality of actions. Therefore, a greedy strategy that always exploits the Q-table and chooses the optimal action from the Q-table may not provide enough exploration for the UAV to guarantee an optimal performance.
A very common approach is to add some randomness to the policy \cite{Singh2000}. We use $\epsilon$-greedy with a decaying exploration, in which a UAV chooses a random exploratory action at state $s$ with probability $\epsilon(s)=c/n(s)$, where $0<c<1$ and $n(s)$ is the number of times the state $s$ has been observed so far. The UAV exploits greedily from its Q-table with probability of $1-\epsilon(s)$. In this approach, the probability of exploration decays over time as the UAVs learn more.

Similar to the original Q-learning for a single agent environments, the proposed learning algorithm converges if the state-action pairs are observed infinitely many times. 
Also, the time complexity 
of 
the algorithm is in 
the order of 
$O(|\mathcal{S}|\times |A^i|)$, where $|\mathcal{S}|$ is the size of the state space, 
and $|A^i|$ is 
the size of action space for UAV $i$. Since there are only two possible actions in our application, the complexity can be expressed as $O(|\mathcal{S}|)$. In terms of space complexity, each UAV $i$ needs to keep a table of size $|\mathcal{S}|\times |A^i|$.

\section{Simulation Results} \label{sec:Simulations}
In this section, we present the simulation results to evaluate the performance of the proposed method. We simulate our system model for 
a ground-based primary transmitter-receiver pair along with the pair of source and fusion for the UAV network. 
The location of primary users, source and fusion are fixed during the simulation. However, the UAVs are distributed randomly in the environment. The channels between nodes $i$ and $j$ are obtained from $h_{i,j} \sim CN(0,d^{-2}_{i,j})$, where $d_{i,j}$ is the distance between nodes $i$ and $j$. The duration of one time slot, $T$, is assumed to be equal to 1. The values of $\gamma_{1}$ , $\gamma_{2}$ and $\gamma_{3}$ are set to $2$, $2$ and $0.4$, respectively. 
\vspace{-3mm}
\subsection*{Scenario I: 2 UAVs}
\begin{figure}[t]
	\centering
	\includegraphics[width=.92\columnwidth]{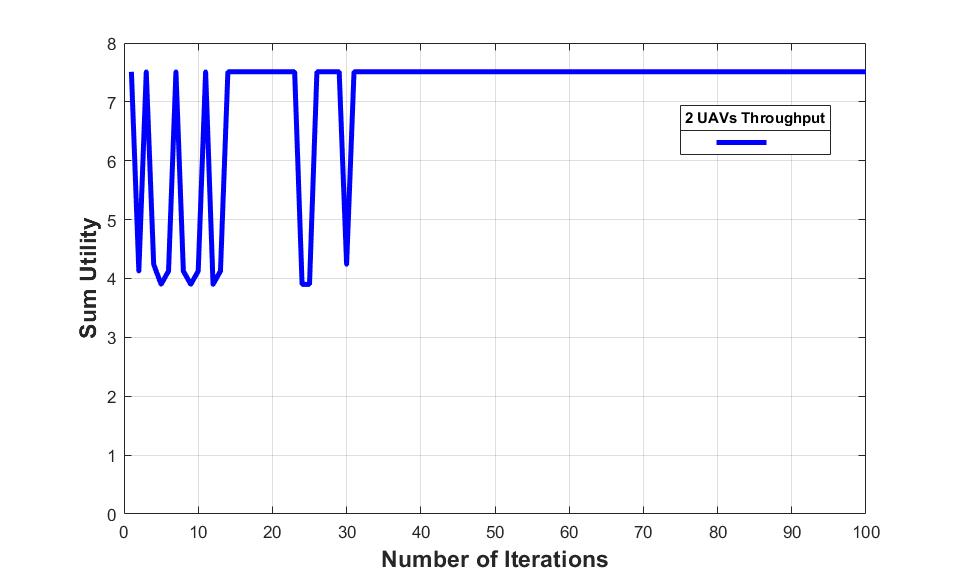}
	\caption{\footnotesize Sum Rate for 2 UAVs for 100 iterations}
		\vspace{-10pt}
    \label{fig:2nodesRate}
\end{figure}

In the first scenario, we consider two UAVs to be partitioned into two task groups. 
\begin{figure}[t]
	\centering
	\includegraphics[width=.92\columnwidth]{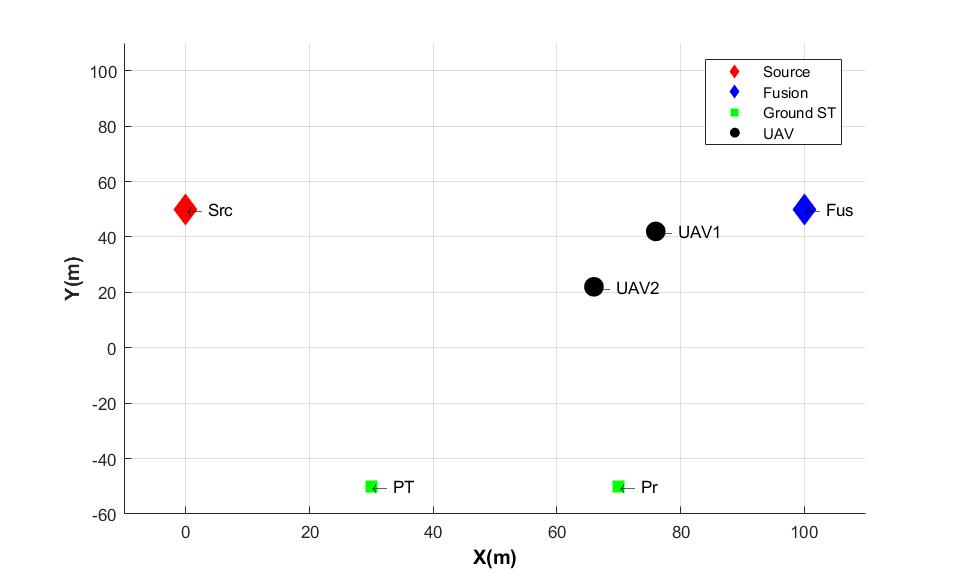}
	\caption{\footnotesize Topology for 2 UAVs in a 100 x 100 mission area.} 
		\vspace{-10pt}
    \label{fig:2nodesSetting}
\end{figure}
The network topology for this scenario is demonstrated in Fig. \ref{fig:2nodesSetting}. Since in this scenario we only have 2 nodes, the possible states for task allocation is equal to $2^{2} = 4$. Hence, the Q-tables will be learned after a few iterations. Fig. \ref{fig:2nodesRate} illustrates the summation of the obtained throughput. 
The convergence to the optimal task allocation 
occurs after 
the 35$^{th}$ iteration, since the number of states is relatively small.
The matrix below shows the final task allocation values for these UAVs.
 \[
\begin{bmatrix} \label{matrix:2nodes}
   0 & 1
\end{bmatrix}
\]
In this notation, 0 corresponds to the set of source-fusion and 1 means the set of the primary users. UAV1 who has a lower relative distance to the source-fusion, is allocated to the fusion set, while UAV2 is allocated to the another set to relay for the primary network. 
\vspace{-3mm}
\subsection*{Scenario II: 6 UAVs}
In this scenario, we consider 6 UAVs to show that the convergence of the proposed method is achieved 
after more iterations compared to the case of 2 UAVs in the first scenario, since the number of states with 6 nodes is equal to $2^6 = 64$. This means, at least 64 iterations are required for the algorithm to just test all the states. 

\begin{figure}[t]
	\centering
	\includegraphics[width=.92\columnwidth]{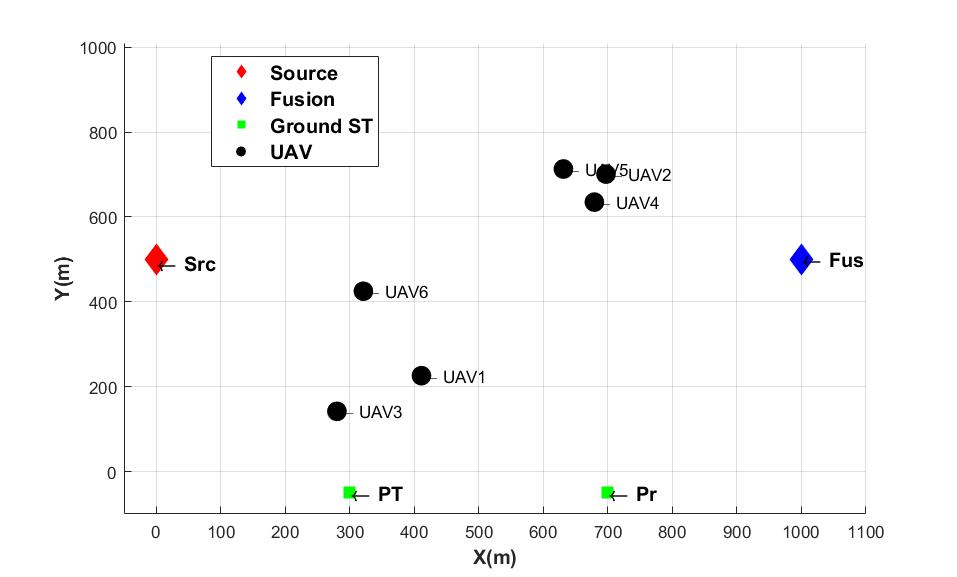}
	\caption{\footnotesize Topology for 6 UAVs in 100 x 100 simulation field}
		\vspace{-10pt}
    \label{fig:6nodesSetting}
\end{figure}

\begin{figure}[t]
	\centering
	\includegraphics[width=.92\columnwidth]{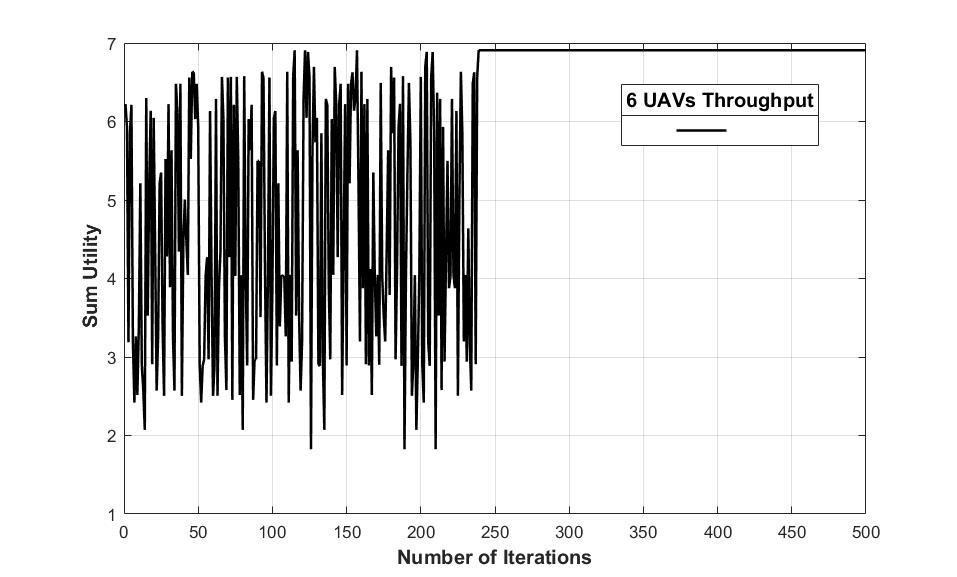}
	\caption{\footnotesize Sum Rate for 6 UAVs for 1000 Iterations}
	\vspace{-10pt}
    \label{fig:6nodesRate}
\end{figure}

\begin{figure}[t]
	\centering
	\includegraphics[width=.92\columnwidth]{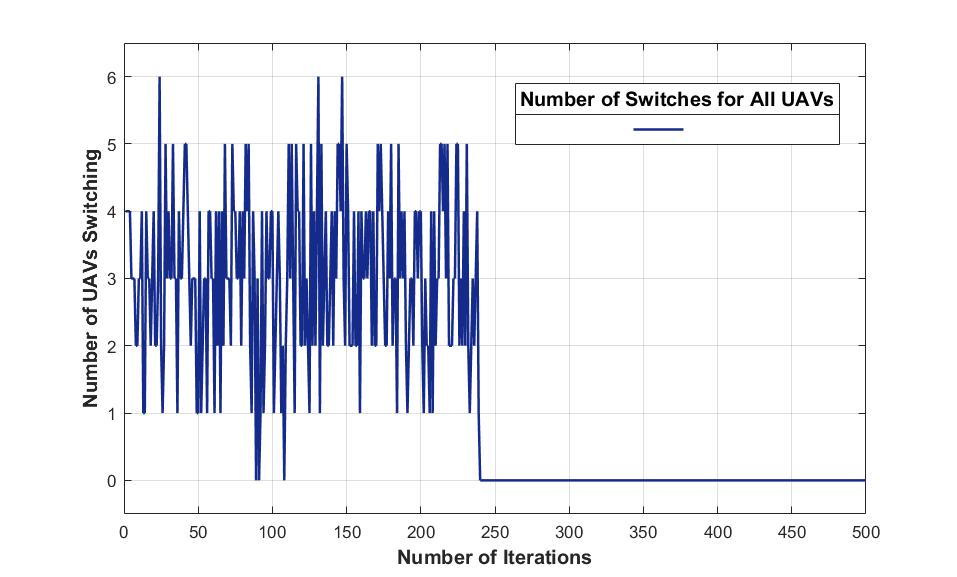}
	\caption{\footnotesize Number of Switching UAVs for 1000 iterations}
		\vspace{-10pt}
    \label{fig:6nodesswitch}
\end{figure}

Fig. \ref{fig:6nodesSetting} demonstrates the network topology with these 6 UAVs
for the primary user and the fusion. As we can see in Fig. \ref{fig:6nodesRate}, the convergence to 
the 
best task allocation occurred 
after 
240 iterations. This implies that the more UAVs 
are 
added to the model, the more iterations will be taken to the convergence
epoch. 
Moreover, Fig. \ref{fig:6nodesswitch} shows the number of UAVs switching their actions (i.e. task partitions) in this scenario. After the 240$^{th}$ iteration, when the convergence happens, we see that no UAV changes its task partition, and the number of switches stays at zero.  

Also, task matrix shown below denotes the final task allocation for the 6 UAVs.
 \[
\begin{bmatrix} \label{matrix:6nodes}
   	 1 & 0 & 1 & 0 & 0 & 1
\end{bmatrix}
\]
Based on this matrix, $\textnormal{UAV}_i$; $i \in \{2,4,5\}$ are considered for the set of source-fusion and 
the rest of UAVs 
are assigned to the relay task group for the primary network. This allocation makes sense considering the location of UAVs and their relative distances.

\color{black}







\section{Conclusion}\label{sec:Conclusion}
In this paper, we studied the task allocation problem for spectrum management in UAV networks. We considered a cooperative relay system in which a group of UAVs 
provide relaying service 
for a 
ground-based 
primary user in exchange for spectrum access. The borrowed spectrum is not necessarily used by the relay UAV, rather 
is used by other UAVs 
to transmit their own information 
to a fusion center. 
This makes a win-win situation for both networks. 
We defined utilities for both the UAV network and the 
ground-based primary network 
based on the achieved rates. 
Next, we proposed a distributed learning algorithm by which the UAVs take proper decisions by joining the relaying or fusion task groups without the need for information exchange or knowledge about other UAV's decisions. 
The algorithm converges to the optimal task partitioning that maximizes the total utility of the system. Simulation results were presented in different scenarios to verify the convergence of the proposed algorithm. 




\bibliographystyle{elsarticle-harv}

\bibliography{sample}

\end{document}